# Reduction of interfacial thermal resistance of overlapped graphene by bonding carbon chains

Yuwen Huang(黄钰文)[1,2], Wentao Feng(冯文韬)[1,2], Xiaoxiang Yu(余晓翔)[1,2], Chengcheng Deng(邓程程)[1*], Nuo Yang(杨诺)[1,2*]

1 School of Energy and Power Engineering, Huazhong University of Science and Technology, Wuhan 430074, P. R. China
2 State Key Laboratory of Coal Combustion, Huazhong University of Science and Technology, Wuhan 430074, P. R. China

* To whom correspondence should be addressed. E-mail: nuo@hust.edu.cn (N.Y.), dengcc@hust.edu.cn (C.D.)

**Abstract**

Exploring the mechanism of interfacial thermal transport and reducing the interfacial thermal resistance is of great importance for thermal management and modulation. Herein, the interfacial thermal resistance between overlapped graphene nanoribbons is largely reduced by adding bonded carbon chains by performing molecular dynamics simulations. And the analytical model (cross-interface model, CIM) is utilized to analyze and explain the two-dimensional thermal transport mechanism at cross-interface. An order of magnitude reduction in interfacial thermal resistance is found as the graphene nanoribbons are bonded by just one carbon chain. Interestingly, the decreasing rate of interfacial thermal resistance slows down gradually with the increasing of the number of carbon chains, which can be explained by the proposed theoretical relationship based on CIM. Moreover, by the comparison of CIM and traditional simplified model, the accuracy of CIM is verified and demonstrated in overlapped graphene nanoribbons. This work provides a new way to improve the interfacial thermal transport and reveal the essential mechanism for low-dimensional materials applied in thermal management.

**Keywords:** phonon engineering, graphene, cross-interface model, molecular dynamics

# 1.Introduction

Nowadays, low-dimensional materials have aroused widespread research interest and show great application prospects.[1-4] Representatively, graphene, a two-dimensional nanomaterial composed of carbon atoms with *sp²* hybrid orbital hexagonal honeycomb lattice, has maintained a strong research interest due to its many intriguing characteristics, such as superior thermal conductivity,[5-9] strong mechanical strength,[10,11] electronic properties[12-16] and optical transparency.[17-19] In particular, due to the superior thermal conductivity (exceeding $\sim 3000\, W/K \cdot m$ near room temperature[6]), graphene plays a growing significant role in thermal management of the next generation electronic devices.[20-25] As a two-dimensional material, graphene has a remarkable characteristic, that is anisotropic heat flow existing in two directions: in-plane and out-of-plane. Covalent $sp^2$ bonding between carbon atoms leads to high in-plane thermal conductivity, whereas out-of-plane heat flow is limited by weak van der Waals coupling.[26]

As an excellent material with high in-plane thermal conductivity, graphene is usually applied in various ways for heat dissipation.[27,28] For example, graphene is used as a filler to synthesize organic compounds into polymer, which optimizes the thermal properties of the organic compounds and serves as a thermal interface material.[29-32] Besides, it is arranged in an orderly stack to synthesize macro-scale films with high thermal conductivity for use as a heat dissipation substrate.[33,34] In practical applications, however, these methods cannot greatly increase the thermal conductivity of the materials. The reason is that there are lots of interfaces between the organic substance and graphene or between graphene and graphene.[35,36] These interfaces usually have large thermal resistance, which significantly affects the thermal transport in the composite system.[37] Therefore, it is necessary to investigate the interfacial thermal transport of graphene-based composites.

There are a large number of nano-scale interfaces observed in composites and films.[38-41] The low-dimensional materials, like graphene,[42] carbon nanotube (CNT), and boron nitride[39] (BN), are staggered in parallel and formed many overlapped interfaces, which are called as cross-interfaces. Different from the unidirectional heat conduction phenomenon at the traditional interface, when the heat flows through this interface, it is simultaneously transported inside and between the ribbons, forming a two-dimensional heat conduction process.[43-46] To reduce the thermal resistance of cross-interface of low-dimensional materials, several methods have been adopted in experimental and simulation works.[47-49] For instance, Qiu et al.[50] promoted interfacial thermal transport by using the intriguing CNT fibers decorated with Au nanoparticles as a promising platform. Liu et al.[51] found that the junctions formed by $-C_2H_4-$ molecular groups between two graphene nanoribbons (GNRs) are effective for transmitting the out-of-plane phonon modes of GNRs. In Xu et al.'s work,[52] polymer wrapping significantly improved the interfacial thermal conductivity of the CNT interfaces. The polymer chains were lying along the groove between CNTs when the wrapping density was assured and assisted the heat transfer process. In this work, a new and simple method is proposed to reduce the interfacial thermal resistance by adding carbon chains bonded between the overlapped graphene nanoribbons.

In previous works, thermal transport at cross-interface was traditionally treated by point contact or by approximation methods. For instance, Yang et al. treated the total thermal resistance of the cross-interface as the sum of the interfacial thermal resistances and the in-plane thermal resistance with half of the overlapping length in series[53,54] (i.e. Simplified Model, SM). Zhong et al. calculated interfacial thermal resistance by treating the overlapped region as a single planar interface between coaxial hot and cold nanotubes joined end to end.[55] Liu et al. calculated the total thermal conductance of overlapped region based on the Fourier law, where the temperature difference was given by the average temperature difference.[51] In experimental works, the total thermal resistance of cross-interfacial sample was usually firstly measured by using a thermal bridge method,[56,57] then one segment was removed and the other was realigned to bridge the two membranes. From these two measurements, the interfacial thermal resistance could be extracted.[54] Generally, these methods treated the interface system approximately as one-dimensional thermal transport process and lacked the accurate mechanism exploration of thermal transport at the cross-interface. Fortunately, an analytical model, named as the cross-interface model (CIM), had been proposed to explore the mechanism of the two-dimensional thermal transport at the cross-interface.[58,59]

In this work, the interfacial thermal resistance at the cross-interface between two overlapped graphene nanoribbons is effectively reduced through bonding carbon chains (CCs), and the theoretical model of CIM is applied to analyze and explain the physical mechanism of thermal transport at the cross-interface. Firstly, the basic idea and framework of CIM theoretical model are presented and molecular dynamics simulation is carried out to calculate the interfacial thermal resistance of two overlapped graphene nanoribbons. Then, the CIM is applied to analyze and explain the interfacial thermal resistances of overlapped graphene nanoribbons, and the influence of CCs' number on the thermal transport is investigated. Lastly, the advantage of CIM relative to the traditional SM for thermal transport at the cross-interface is compared and discussed. Our study not only provides a new way to improve the interfacial thermal transport of graphene-based composites, but also deepens the understanding of the thermal transport at the cross-interfaces.

## 2.Theoretical model and simulation method

**2.1 Cross-interface model**

The cross-interface model (CIM) is utilized to calculate and analyze the thermal transport in overlapped graphene nanoribbons. Here, CIM is derived on the base of Boltzmann transport equation (BTE). There is size effect in micro/nanoscale phonon transport when the size of structures comparable to the mean free path of phonons. For classical size effects which do not consider quantum effect, the phonon transport is largely affected by boundary scatterings. Transport processes of classical size effects can be treated on the basis of BTE by treating phonons as particles.[60] Moreover, from Ma et. al.'s work, it can be seen that the particle characteristics of phonon transport still exist widely in the nanoscale materials.[61] Therefore, It is worth to note that BTE is not only applicable to diffusive transport, but also suitable for ballistic transport under certain conditions.[60]

For such a cross-interface, the heat flows along the structure horizontally and through the interface vertically simultaneously. The CIM is applicable to illustrate the two-dimensional thermal transport process at the cross-interface by considering the coupling effect between different thermal transport channels. The coupling effects may be much weaker than phonon transport in each ribbon, but it is not negligible for the thermal transport in the entire system. Referring to the works of Feng et al.[59] and Xiong et al.[58], the CIM was based on Fourier's law and energy conservation. In the overlapped region, the heat conduction equations for each ribbon are given below.

$$\kappa \frac{d^2 T_T}{dx^2} A - G_{CA}(T_T - T_B)w = 0, 0 < x < L_C \tag{1a}$$

$$\kappa \frac{d^2 T_B}{dx^2} A + G_{CA}(T_T - T_B)w = 0, 0 < x < L_C \tag{1b}$$

where $T_T$ and $T_B$ are the temperatures of the top and bottle GNRs, respectively. $w$ is the width of GNRs and $A$ is cross sectional area. $G_{CA}$ is the interfacial thermal conductance per unit area and $\kappa$ is the thermal conductivity of the two same GNRs. $L_C$ is the length of the overlapped region.

Through a series of formula conversions,[59] the temperature distribution functions can be represented as follow:

$$T_T = \frac{1}{2}(a \times e^{-\gamma x} + b \times e^{\gamma x} + cx + d) \tag{2a}$$

$$T_B = \frac{1}{2}(-a \times e^{-\gamma x} - b \times e^{\gamma x} + cx + d) \tag{2b}$$

$$\gamma = \sqrt{2 G_{CA} w / \kappa A} \tag{3}$$

where $a, b, c, d$ are parameters and the details are showed in supporting information.

The thermal resistance can be deduced as follow:

$$R_{total} = R_{intra} + \frac{1}{\eta} \times R_{inter} \tag{4}$$

$$\eta = \frac{tanh\sqrt{R_{intra}/R_{inter}}}{\sqrt{R_{intra}/R_{inter}}} \tag{5}$$

where $R_{total}$ is the total thermal resistance; $R_{intra}$ is the intra-ribbon thermal resistance assuming that the top ribbon and the bottom ribbon transport heat flux in a parallel model, just like the parallel law in electrical transport; $R_{inter}$ is the inter-ribbon thermal resistance, which is the inverse of interfacial thermal conductance. And $\eta$ is a factor related to the ratio of $R_{intra}$ and $R_{inter}$. The detailed explanation of the formula is shown in supporting information.

## 2.2 MD simulation method

In this work, the nonequilibrium molecular dynamics (NEMD) is utilized to calculate the interfacial thermal resistance between two overlapped GNRs.[62-64] The simulation system consists of two same GNRs, which are overlapped in z direction as shown in Fig. 1. Each GNR is 10.19 nm long and 3.05 nm wide, and the length of overlapped region ($L_C$) is 9.21nm. The interlayer distance between the two GNRs is

0.35nm. Different numbers of carbon chains (CC), including 1, 3, 5, 7, are respectively bonded between two GNRs cross the overlapped section. The carbon atoms in the CC are connected alternately by single and double bonds; the CC and GNR are bonded by $sp^3$ bonds. The three bonds on each atom of CC are distributed in the same plane and the bond angles are all 120° initially. The CCs are evenly distributed in the width direction of GNR. In order to ensure that the atomic layers at both ends of the overlapped-GNRs are connected by CCs, adjacent CCs are staggered by a C-C bond horizontal distance ($L_o$ in Fig. 1d) in the z direction.

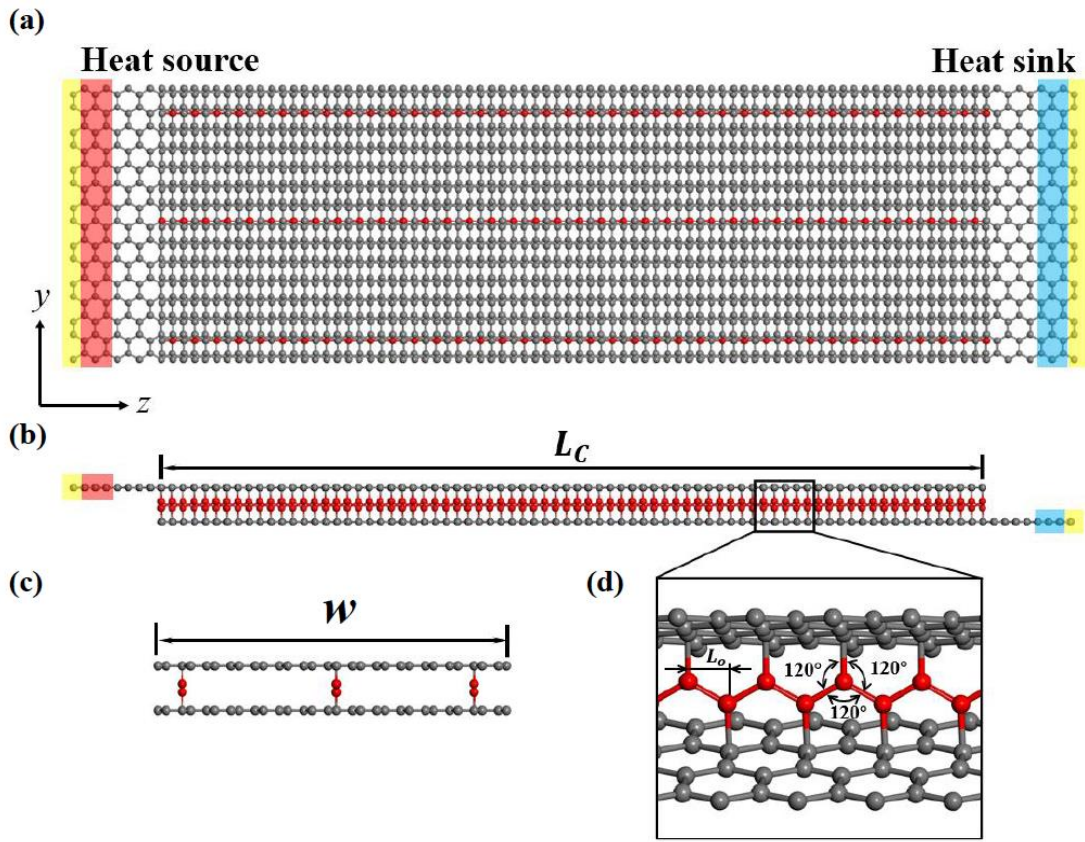

**Fig. 1.** Schematic of two overlapped graphene nanoribbons bonded by three carbon chains: (a) top view; (b) front view; (c) side view; (d) the local enlarged drawing of bonded CCs (red chains) between the two GNRs.

In the MD simulations herein, the Tersoff potential is employed to describe the interatomic interactions, while it has successfully tested the thermal conductivity of

GNR[63,65] and the thermal properties of graphene.[62,66,67] The van der Waals interactions between two GNRs are described by Lenard-Jones (L-J) potential,[52] $V_{ij} = 4\varepsilon[(\sigma/r)^{12} - (\sigma/r)^6]$, $\varepsilon = 0.002968 eV$, $\sigma = 3.407 Å$, where $\varepsilon$ is the depth of the potential energy well, $\sigma$ is the finite distance at which the interatomic potential is zero, and $r$ is the distance between the atoms. The time step of all simulations is set as 0.25 fs, and the velocity Verlet algorithm is used to integrate the discrete differential equations of motion.[68]

All the simulations here are performed by the large-scale atomic/molecular massively parallel simulator (LAMMPS) packages,[69] which has been widely used to study thermal transport properties in nanoscale.[70-72] In the simulation structure, the size of the simulation box is $10 \times 10 \times 19.65$ nm$^3$, which is much larger than the initial volume of overlapped graphene. The GNR structure was firstly equilibrated at a constant temperature for 250 ps in the canonical ensemble (*NVT*) and then in microcanonical ensemble (*NVE*) for 100 ps. Then a layer of atoms at both ends of the structure is fixed (yellow region in Fig. 1). Afterwards, a heat source with a higher temperature of 320 K is applied to the atoms in the red region and the heat sink with a lower temperature of 280 K is applied to atoms in the blue region using Langevin thermostats. After the system reached steady state by performing simulation for 2.5ns, a time averaging of temperature and heat current was performed for an additional 25 ns to get the temperature profile and the value of heat flux (NEMD simulation details are given in supplementary material). And the reliability of the simulation method is verified by calculating the thermal conductivity of the GNR sheets and the thermal conductivity of the GNR is consistent with previous works[73] (related calculation details shown in supplementary material).

## 3.Results and discussion

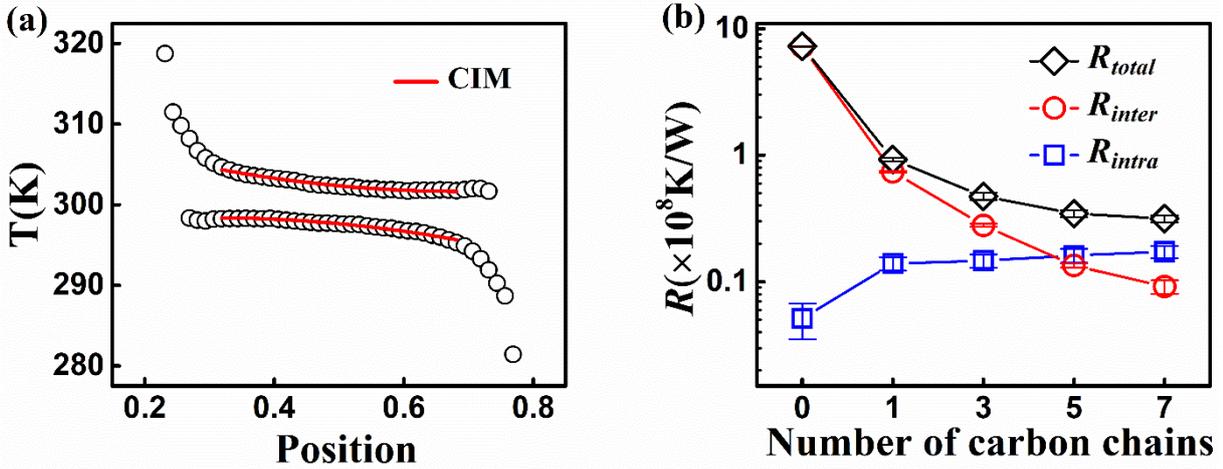

**Fig. 2.** (a) Temperature profile of overlapped graphene nanoribbons bonded by three CCs. The expression of temperature distribution using CIM is fitted to the simulation result as the red line. (b) Influence of the number of CCs on thermal resistance at the cross-interface.

Firstly, NEMD method is performed to obtain temperature profile of the structures, then the Eq. (1) in the CIM theoretical model is used to fit the calculation results in the overlapped section. A typical temperature profile of the interface structure bonded by three CCs is presented in Fig. 2(a). It is seen that the results are fitted fairly well by using the CIM (the red line). And the coefficient of determination ($R^2$) is equal to 0.99829, which exhibits good quality of the fitting.

Three kinds of thermal resistances are obtained through the calculations. As shown in Fig. 2(b), with the number of CCs increasing, the total thermal resistance ($R_{total}$) and inter-ribbon thermal resistance ($R_{inter}$) decreases significantly, and the intra-ribbon thermal resistance ($R_{intra}$) increases slowly. When no CC is bonded, interfacial thermal resistance is very large because the van der Waals interaction between the two graphene ribbons is extremely weak. The $R_{total}$ and $R_{inter}$ for the case with no CC are equal to $7.27 \times 10^8 \, K/W$ and $7.20 \times 10^8 \, K/W$, respectively. After one CC is bonded, the value of $R_{total}$ and $R_{inter}$ are $9.25 \times 10^7 \, K/W$ and

$7.40 \times 10^7 \, K/W$, respectively, which are nearly one order of magnitude lower than the case with no CC. It is confirmed that covalent bond interactions dominate in the interfacial thermal transport which is much stronger than van der Waals interactions. It is noteworthy that when the number of CCs increases to three, the thermal resistance tends to converge. This reason will be discussed later. Moreover, the values of $R_{intra}$ basically keep the same when the number of CCs increases from one. This indicates that the thermal conductivity of GNR remains unchanged with the increasing number of CC bonded in the out-plane direction.

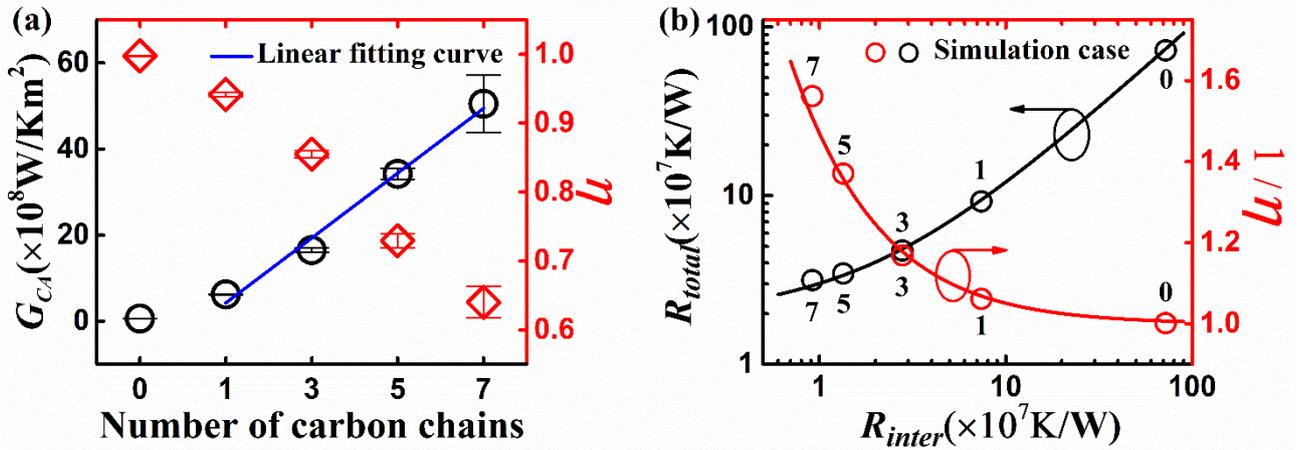

**Fig. 3** (a) Interfacial thermal conductance per unit area (black hollow dot) and value of dimensionless factor $\eta$ (red hollow diamond) as a function of the number of CCs. Blue line is the linear fitting line from NEMD results. (b) $R_{total}$ (black line) and $1/\eta$ (red line) as a function of $R_{inter}$. The hollow dots show the five simulation results, and the Arabic numerals represent the number of CCs.

The variation of interfacial thermal conductance per unit area ($G_{CA}$) with the number of CCs is depicted in Fig. 3(a). When no CC is bonded, the value of $G_{CA}$ is $6.39 \times 10^7 \, W/K \cdot m^2$. After one CC is bonded, the value of that rises to $6.22 \times 10^8 \, W/K \cdot m^2$, which is nearly one order of magnitude higher. It is confirmed again that covalent bond interactions greatly enhance the interfacial thermal conductance. More interestingly, the interfacial thermal conductance per unit area nearly follows a linear relationship with the number of CCs (shown as blue line in Fig. 3(a)). This indicates that each CC serves as an independent channel for heat conduction, and there is little coupling between adjacent CCs.

It can be seen from Fig. 2(b) that the decreasing rate of $R_{inter}$ slows down gradually with the increasing of the number of boned CCs. In order to analyze this tendency theoretically, the $G_{CA}$ is considered to contain two parts: one is devoted by van der Waals interactions and the other is contributed by covalent bond interactions. Based on these assumptions and analysis, interfacial thermal conductance per unit area can be given by

$$G_{CA} = N \cdot \Delta G + G_0 \tag{6}$$

where $\Delta G$ represents interfacial thermal conductance per unit area that each CC possesses and $N$ represents the number of CCs. Here, when $N = 0$ for the case without bond-chain, $G_{CA} = G_0$. Therefore, $G_0$ represents the value of interfacial thermal conductance per unit area in the case without bond-chain.

According to Eq. (6) and Eq. (S4c), $R_{inter}$ can be derived as

$$R_{inter} = \frac{1}{wL_C \cdot \sqrt{(N \cdot \Delta G + G_0)}} \tag{7}$$

It can be noted that there is a roughly proportional relationship ($R_{inter} \propto 1/\sqrt{N}$) in Eq. (7). Therefore, when $N$ further increases, the decreasing rate of $R_{inter}$ slows down and the $R_{inter}$ tends to numerical convergence.

Furthermore, the relationship curve of $R_{total}$ and $R_{inter}$ is given in Fig. 3(b). Herein, the $R_{intra}$ is regarded as constant in Eq. (4) because it represents the intrinsic thermal resistance of the GNR and the calculation results in Fig. 2(b) basically keep the same when the number of CCs increases from one. The results show that when CCs' number increases ($R_{inter}$ decreases), the $R_{total}$ decreases monotonically. Since $\eta$ is a coefficient related to the $R_{inter}$ according to Eq. (5), the relationship curve of $1/\eta$ and $R_{inter}$ is shown as red line in Fig. 3(b). Although the $1/\eta$ increases to infinity when the $R_{inter}$ decreases, it cannot change the monotonically decreasing trend of $R_{total}$. This can be theoretically explained that increasing the number of CCs in the overlapped area significantly reduces the thermal resistance of the overlapped graphene. Moreover, the simulation value points of five cases are depicted as hollow dots in Fig. 3(b). These simulation dots fall well on the theoretical curve, which confirms the accuracy of the CIM.

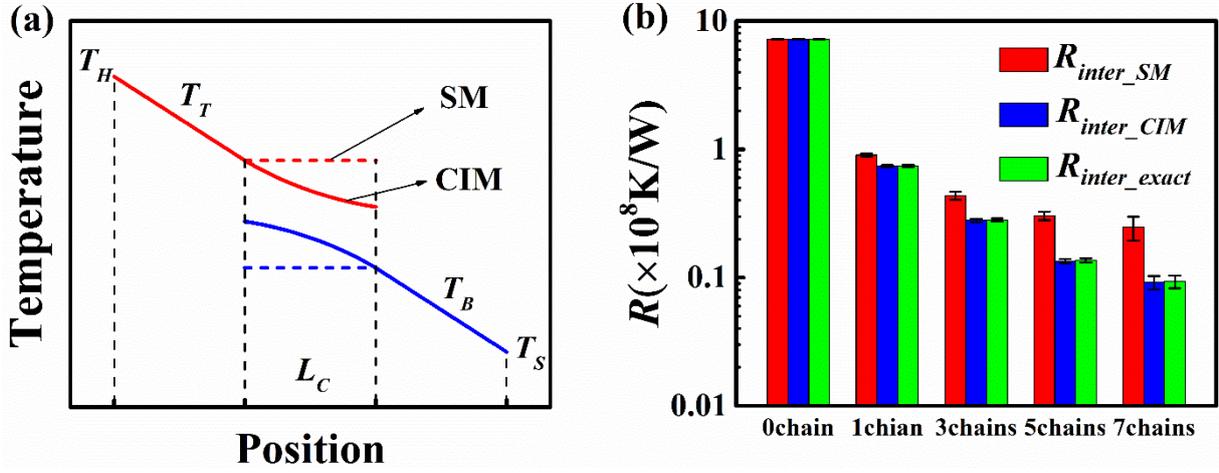

**Fig. 4** (a) Theoretical curve of temperature distribution using CIM and SM. (b) The comparison of the interfacial thermal resistance calculated by SM, CIM and average temperature difference, i.e. $R_{inter\_SM}$ (red bar), $R_{inter\_SIM}$ (blue bar), and $R_{inter\_exact}$ (green bar) respectively, in five cases with different numbers of CCs.

To compare the difference between CIM and traditional simplified model (SM), the theoretical curves of temperature distribution using CIM and SM are shown in Fig. 4(a). The solid exponential curve in the overlapped region is drawn by Eq. (2), which shows the temperature distribution at the cross-interface using CIM. The horizontal dashed line corresponds to the temperature distribution by using SM. It is known that the SM cannot present a specific image of temperature distribution, i.e. black box. In SM, it is considered that the total thermal resistance is a simple series connection of intrinsic thermal resistance and interfacial thermal resistance. As for CIM, it can provide a clear physical image of temperature distribution and reveal the mechanism of the two-dimensional thermal transport at the cross-interface. By using CIM, the total thermal resistance is represented by a series of intrinsic thermal resistance and interfacial thermal resistance with a specific interacting relationship.

In order to assess the accuracy of CIM, the $R_{inter\_SM}$, $R_{inter\_SIM}$, and $R_{inter\_exact}$, which are the interfacial thermal resistance calculated by SM, CIM and average temperature difference method, respectively, are compared in Fig. 4(b). Here, to make the description and comparison of variables more intuitionistic, suffix is added to the subscript of the variable $R_{inter}$ to distinguish the values calculated by these three methods. $R_{inter\_SM}$ denotes the inter-ribbon thermal resistance calculated by simplified model. $R_{inter\_CIM}$ denotes the inter-ribbon thermal resistance calculated by cross-interface model. $R_{inter\_exact}$ denotes the inter-ribbon thermal resistance calculated by the ratio of the temperature difference to the heat flux.

Based on idea of experimental measurements, $R_{inter\_SM}$ is calculated by the difference between $R_{total\_SM}$ and $R_{intra\_SM}$. The formula using the SM can be given as

$$R_{inter\_SM} = R_{total\_SM} - R_{intra\_SM} = \frac{T_T|_{x=0} - T_B|_{x=L_C}}{J} - \frac{L_C}{2\kappa_0 A} \tag{8}$$

where $\kappa_0$ is the thermal conductivity of bare GNR which means no CC is bonded between the GNRs. As for $R_{inter\_exact}$, it is calculated by the ratio of the temperature difference to the heat flux which is recognized as an accurate calculation formula for solving interfacial thermal conductivity[47,51,55]:

$$G_{CA} = \frac{J}{w(\overline{T_T} - \overline{T_B})L_C} \tag{9a}$$

$$R_{inter\_exact} = \frac{1}{G_{CA} w L_C} \tag{9b}$$

where $\overline{T_T}$ and $\overline{T_B}$ denote the average temperature of $T_T$ and $T_B$, respectively. It is noteworthy that although $R_{inter\_CIM}$ and $R_{inter\_exact}$ are both calculated by the idea of average temperature difference, the average temperature in $R_{inter\_CIM}$ is calculated by integral of the fitted function while that in $R_{inter\_exact}$ is obtained through arithmetic mean of temperature data in overlapped region. Thus, a significant advantage of CIM over average temperature difference method is that the theoretical equation of temperature distribution in the overlapped area can be obtained, which will help explore the mechanism of interfacial thermal transport.

From Fig. 4(b), it can be shown that $R_{inter\_SIM}$ always retain almost consistent with $R_{inter\_exact}$, however $R_{inter\_SM}$ appears some deviation compared with $R_{inter\_exact}$, especially when the number of CCs increases more. The reason is that when CC is bonded, the thermal conductivity of GNR in the system is no longer equal to that of bare GNR (i.e. $\kappa_0$). However, based on the idea of SM, the thermal conductivity of bare GNR is still applied to calculate the $R_{inter\_SM}$ and the deviation will be more obvious when more CCs are bonded. For CIM, it can exhibit two-dimensional thermal transport process and obtain clear temperature distribution profile which make it possible to directly calculate $R_{inter\_SIM}$ accurately. Therefore, the accuracy of CIM relative to SM is verified in the example of overlapped GNRs. Furthermore, the CIM can be used to better guide how to reduce the interface thermal resistance of graphene-based composites by adding the bonded carbon chains.

## 4. Conclusion

In summary, the interfacial thermal resistance between overlapped graphene nanoribbons is reduced by bonding carbon chains through NEMD simulation, and the CIM analytical model is applied to analyze and explain the interfacial thermal transport mechanism. After one CC is bonded, the interfacial thermal resistance is reduced by an order of magnitude because covalent bond interactions largely enhance the interfacial thermal transport. While more CCs are bonded, the decreasing rate of interfacial thermal resistance slows down gradually, which can be explained by the proposed numerical relationship between thermal resistance and number of CCs based on CIM, i.e. $R_{inter} \propto 1/\sqrt{N}$. Moreover, the advantage of CIM relative to the traditional simplified model is demonstrated in the example of overlapped GNRs especially when more CCs are bonded. This work can provide valuable guide for the design and application of graphene-based materials for effective thermal management and modulation.

## Conflicts of interest

There are no conflicts to declare.

# Acknowledgements

This work was financially supported by the National Natural Science Foundation of China (No. 51606072 (C. D.)) and the Fundamental Research Funds for the Central Universities, HUST (No. 2019kfyRCPY045). We are grateful to Xiao Wan and Shichen Deng for their useful discussions. The authors thank the National Supercomputing Center in Tianjin (NSCC-TJ) and High Performance Computer Cluster in Huazhong University of Science and Technology (HUST-HPCC) for providing assistance with computations.